\documentstyle[amstex,amssymb,epsf]{lamuphys}

\newtheorem{th}{Theorem}
\newtheorem{pro}{Proposition}

\renewcommand{\I}{\operatorname{i}}

 \newcommand{\sdet}{\operatorname{sdet}}

\newcommand{\C}{{\Bbb C}}

\begin{document}

\title{Darboux Transformations for
SUSY Integrable  Systems}

\author{Q. P. Liu\thanks{On leave of absence from
Beijing Graduate School, CUMT, Beijing 100083, China}
\thanks{Supported by {\em Beca para estancias temporales
de doctores y tecn\'ologos extranjeros en
Espa\~na: SB95-A01722297}}
   $\,$ and Manuel Ma\~nas\thanks{Partially supported by CICYT:
 proyecto PB95--0401}}
\institute{Departamento de F\'\i sica Te\'orica,
\\ Universidad Complutense,\\
E28040-Madrid, Spain.}

\maketitle

\begin{abstract}
Several types of Darboux transformations for supersymmetric
integrable systems
such as the Manin-Radul KdV, Mathieu KdV and SUSY sine-Gordon equations
are considered. We also present solutions such as supersolitons and
superkinks.
\end{abstract}
\newpage

\section{Introduction}
Supersymmetric integrable systems constitute a subject of current
interest, and as a consequence a number of well known integrable
equations have been generalized into the supersymmetric (SUSY) context.
We just mention the SUSY versions of sine-Gordon (\cite{di,c-k,f-g-s}),
 Nonlinear Schr\"odinger (\cite{rk}),
KP (\cite{m-r,rabin,mulase}), KdV (\cite{m-r,ma2})
and Boussinesq (\cite{yung,b}).
 We also point out that there are two different
generalizations, namely nonextended ($N=1$, $N$ being the number
of fermionic independent variables) and extended ($N\geq 2$) generalizations.
Here we are mainly interested in the former case.
So far many of the tools used in the standard theory have been
extended to this framework, such as B\"acklund transformations
 (\cite{c-k}),
prologation theory (\cite{rk}),
Hamiltonian formalism (\cite{op,mp}), Grasmmanian description (\cite{ueno}),
tau function (\cite{ueno,luis}),
the relationship with super $W$ algebras (\cite{ma1,b-g}), 
additional symmetries (\cite{das,mmm,ss}), etc..

However, there is one that only recently has been considered, we
are talking about the so called {\em Darboux transformation}, that
constitutes a very successful tool in the realm of integrable systems
whenever one is interested in the constructions of solutions. The roots
of these techniques go back to geometrical studies of the last
century. It was in \cite{moutard}, see also \cite{a-n}, where the two dimensional
Schr\"odinger equation (as we called it today) was considered
giving new wave functions and potentials from given ones. This was
taken by Darboux (\cite{darboux1}) and applied to the
one-dimensional case. In fact these results are connected with the
theory of conjugate nets (\cite{darboux2,eisenhart}) and it was in
\cite{levy} where a transformation of this type was applied for
surfaces, and was iterated in \cite{hammond}. New transformations
for conjugate nets appeared in \cite{jonas} which were called
fundamental in \cite{eisenhart2} containing the Levy ones in
appropriate limits.

 In \cite{crum}, independently of these geometrical studies, it was
presented, for the Schr\"odinger equation, the iteration of the
transformation found by Darboux, giving compact expressions in
terms of Wro\'nski determinants. Later on, in \cite{wadati}, 
a {\em new} transformation for soliton equations
was introduced. This tool was rapidly devoloped and it was in
\cite{matveev} where they were named as Darboux transformations.
This name is standard nowadays in the soliton community, however we
have seen that is not completely appropriate. In \cite{levi} a
further extension of the Darboux transformation was given and some
people refer to it as binary or Darboux-Levi transformation,
however this is just the fundamental type transformation mentioned
above. Finally, we remark that recently (\cite{gm,m}) a vectorial
formulation of the binary Darboux transformation was given,
allowing compact formulae for the iteration of the Darboux
transformation, for the Nonlinear Schr\"odinger and
Davey-Stewartson equations.

The paper \cite{liu}   was the first one that considered Darboux
transformations for the SUSY KdV system. Later on
(\cite{lm1,lm2}) extensions of the binary Darboux transformation
(fundamental transformation in Geometry) and 
Darboux transformation  appeared. In this paper
we want to present the SUSY version of the Darboux tranformations.
In particular we will consider three important supersymmetric
integrable systems, namely the Manin-Radul KdV and its reduction to
the Mathieu KdV, and also the SUSY extension of the sine-Gordon
equation. For the Manin-Radul KdV and vectorial binary Darboux
transformations we improve the presentation of \cite{lm1} giving
the general transformation for wave functions and a permutability
theorem. For the Mathieu KdV equation we present some technical
improvements with respect to \cite{liu} and the part regarding the
sine-Gordon equation is entirely new. Finally, let us remark that
given the character of these proceedings we are not going to give
any proof.

The layout of the paper is as follows. We start with the
Manin-Radul KdV (MRKdV) by considering the vectorial binary Darboux
transformations and the construction of solutions in terms of
ordinary determinants of Grammian type, we also give the iteration
of the Darboux transformation found in \cite{liu} to get Wro\'nski
superdeterminantal expressions for the solutions,  here we
present a genuine supersoliton. In
\S 3 we study the application of the Darboux transformation for the
Mathieu KdV to the SUSY sine-Gordon equation, presenting a
superkink.

\section{Darboux Transformations for the Manin-Radul KdV Equation}

The MRKdV system is defined in terms of three independent variables
$\vartheta,x,t$, where $\vartheta\in\Bbb C_{\text{a}}$ is an odd
supernumber, and $x,t\in\Bbb C_{\text{c}}$ are even supernumbers,
and two dependent variables $\alpha(\vartheta,x,t),
u(\vartheta,x,t)$, where $\alpha$ is an odd function taking values
in $\Bbb C_{\text{a}}$ and $u$ is even function with values in
$\Bbb C_{\text{c}}$. A basic ingredient is a superderivation
defined by $D:=\partial_{\vartheta}+\vartheta \partial_x$. The
system is
\begin{equation}\label{mr}
\begin{aligned}
\alpha_t&={1\over 4} (\alpha_{xxx}+3(\alpha D\alpha)_x+6(\alpha u)_x),\\
u_t&={1\over 4}(u_{xxx}+6uu_x+3\alpha_xDu+3\alpha (Du_x)),
\end{aligned}
\end{equation}
where we use the notation
$f_x:=\partial f/\partial x$ and $f_t:=\partial f/\partial t$.

The following linear system for the wave function $\psi(\vartheta, x,t)$,
that takes values in the Grassmann algebra $\boldsymbol\Lambda=\Bbb
C_{\text{c}}\oplus\Bbb C_{\text{a}}$,
\begin{equation}\label{linear}
\begin{aligned}
L(\psi)&:=\psi_{xx}+\alpha D\psi +u\psi=\lambda \psi,\\
\psi_t=M(\psi)&:={1\over 2}\alpha(D\psi)_x+\lambda\psi_x+
{1\over2}u\psi_x
-{1\over 4}\alpha_x D\psi-{1\over 4}u_x\psi,
\end{aligned}
\end{equation}
where the spectral parameter $\lambda\in\Bbb C_{\text{c}}$ is an
even supernumber, has as its compatibility condition Eqs.
(\ref{mr}), and therefore it can be considered as a Lax pair for
(\ref{mr}).

\subsection{Vectorial Binary Darboux Transformations}

The linear system (\ref{linear}) is of a scalar nature,
$\lambda\in\Bbb C_{\text{c}}$,
$\psi(\vartheta,x,t)\in\boldsymbol\Lambda$.
 Nevertheless, it is possible to give a vector extension
of these linear problem.
Indeed, we may replace $\boldsymbol \Lambda$
by an arbitrary linear Grassmann space
$\cal E$ over $\boldsymbol \Lambda$
and take $b$ as an $\cal E$-valued eigenfunction, then the spectral parameter
can be taken as  $\ell\in  \;\text{L}(\cal E_{\underline{0}})\oplus
\text{L}({\cal E}_{\underline{1}})$,  an even operator.

Namely, the linear system
\begin{equation}\label{veclin}
\begin{aligned}
 \psi_{xx}+\alpha D\psi+u\psi-\ell\psi&=0,\\
 \psi_t-{1\over2} \alpha (D\psi_x) -\ell\psi_x-{1\over 2}u\psi_x+
{1\over 4}\alpha_x D\psi
+{1\over 4}u_x\psi&=0,
\end{aligned}
\end{equation}
has as its compatibility condition the
MRKdV system (\ref{mr}).

Notice  that Eqs. (\ref{mr}) are also the compatibility
condition of adjoint linear system:
\begin{equation}\label{aveclin}
\begin{aligned}
\phi_{xx}+D(\alpha \phi )+u\phi -\phi m&=0,\\
\phi_t+{1\over 2}\alpha D\phi_x-\phi_x m-{1\over 2}
(u+D\alpha )\phi_x +{1\over 4}D(\alpha_x\phi)+{1\over 4}u_x\phi&=0,
\end{aligned}
\end{equation}
where $\phi(\vartheta,x,t)\in\tilde{{\cal E}}^*$ is a linear function on the
supervector space $\tilde{{\cal E}}$, and
$m \in \text{L}(\tilde{{\cal E}}_{\underline{0}})\oplus
\text{L}(\tilde{{\cal E}}_{\underline{1}})$.

In order to construct Darboux transformations for these linear
systems we need to introduce an  operator, say
$V[\psi,\phi]\in\text{L}({\cal E},\tilde{\cal  E})$, bilinear in
$\psi$ and $\phi$, defined by the compatible equations
\begin{equation}\label{pont}
\begin{aligned}
DV[\psi,\phi]=&\psi\otimes \phi, \\
V[\psi,\phi]_t=&\ell V[\psi,\phi]_x +V[\psi,\phi]_xm \\&-
D(\psi_x\otimes\phi_x+{1\over 2}uDV[\psi,\phi])
-{1\over 4}\alpha_x DV[\psi,\phi]\\
&-{1\over2}(D\psi)\otimes((D\alpha)\phi-\alpha
(D\psi))+{1\over2}\alpha(\psi\otimes \phi_x -\psi_x\otimes \phi )
\end{aligned}
\end{equation}
such that
\begin{equation}\label{constraint}
\ell V[\psi,\phi]-V[\psi,\phi]m =D(\psi_x\otimes\phi-\psi\otimes \phi_x)-
\alpha \psi\otimes \phi.
\end{equation}

Now we are ready to present the following:

\begin{th}
Let $\psi_0(\vartheta,x,t)\in\cal V_{\underline{0}}$ be an even
vector satisfying Eq. (\ref{veclin}) with spectral parameter $\ell_0$,
$\phi_0(\vartheta,x,t)\in\cal V_{\underline{0}}^*$  an odd functional
solving Eq. (\ref{aveclin}) with spectral parameter $m_0$
and $V[\psi_0,\phi_0]\in \text{L}(\cal
V_{\underline{0}})\oplus\text{L}(\cal V_{\underline{1}})$  a non
singular even
operator, $\det V[\psi_0,\phi_0]_{\text{body}}\neq 0$, defined in terms
of the compatible Eqs.
(\ref{pont}) and (\ref{constraint}).
Then, the objects
\begin{align*}
&\hat{\psi}:=\psi-V[\psi,\phi_0]V[\psi_0,\phi_0]^{-1}\psi_0,\\
&\hat{\phi}:=\phi-\phi_0V[\psi_0,\phi_0]^{-1}V[\psi_0,\phi],\\
&\hat\alpha=\alpha-2D^3\ln\det V[\psi_0,\phi_0],\\
&\hat u=u+2\hat\alpha D\ln\det V[\psi_0,\phi_0]+2\left(
\frac{\sum_{j}D(\psi_0)_j\;\det V[\psi_0,\phi_0]_j}{\det V[\psi_0,\phi_0]}\right)_x,
\end{align*}
where  $V[\psi_0,\phi_0]_j$ is an operator with associated
supermatrix obtained from the corresponding one of
$V[\psi_0,\phi_0]$ by replacing the $j$-th column by $\psi_0$,
satisfy the Eqs. (\ref{veclin}) and (\ref{aveclin}) whenever the
unhatted variables do. Thus, $\hat\alpha$ and $\hat u$ are new
solutions of the MRKdV (\ref{mr}). Moreover,
\begin{equation}\label{newv}
V[\hat\psi,\hat\phi]=V[\psi,\phi]-
V[\psi,\phi_0]V[\psi_0,\phi_0]^{-1}V[\psi_0,\phi].
\end{equation}
\end{th}

Let us remark that this theorem extends Theorem 1 in our paper
(\cite{lm1}). In particular we stress the role of the general wave
functions $\psi$ and $\phi$, and also the formula (\ref{newv}) that
gives the path for iteration and it is also  deeply connected, in
the non SUSY case, with geometrical objects such as points of the
transformed manifolds.

We shall call $({\cal V},\psi_0,\phi_0)$ as transformation data.
The composition of two vectorial Darboux transformations yield a
new Darboux transformation, and as it is shown in next proposition
they commute as they can be expressed as a vectorial Darboux
transformation:
\begin{pro}
The vectorial Darboux transformation with transformation data
$({\cal V}_1\oplus{\cal V}_2,
\Big(\begin{smallmatrix} \psi_{0,(1)}\\ \psi_{0,(2)}\end{smallmatrix}\Big),
(\phi_{0,(1)},\phi_{0,(2)}))$ coincides with the following
composition of Darboux transformations:
\begin{enumerate}
\item First transform with data $({\cal V}_2,\psi_{0,(2)},\phi_{0,(2)})$,
 and denote the transformation by $^\prime$.
\item On the result of this transformation apply a second one with data
   \[
(\cal V_1, \psi_{0,(1)}^\prime, \phi_{0,(1)}^\prime).
\]
\end{enumerate}
\end{pro}
A similar theorem in a completely different framework, namely
discrete integrable systems: multidimensional quadrilateral
lattices, appears in \cite{dms}.

\subsection{Wro\'nski Superdeterminants Representation
of Iterated Darboux Transformations}

A Darboux transformation for the MRKdV equation is (\cite{liu})
\begin{pro} Let $\psi$ be a solution of (\ref{linear})
and $\theta_0$ be a particular solution with  $\lambda=\lambda_0$.
Then, the quantities defined by
\begin{align*}
{\hat{\psi}}: &=(D+\delta_0)\psi,
 \qquad \delta_0:=-{D\theta_0\over \theta_0}, \quad ({\theta_0}: \text{even})\\
\hat{\alpha}:&= -\alpha -2\partial\delta_0,\\
\hat{u}: &=u+(D\alpha)+2\delta_0(\alpha+\partial\delta_0)
\end{align*}
satisfy
\begin{align*}
\hat L\hat\psi&=\lambda\hat\psi,\\
\partial_t\hat \psi&=\hat M\hat\psi,
\end{align*}
where $\hat L$ and $\hat M$ are obtained from $L$ and $M$ by replacing
$\alpha$ and $u$ with $\hat\alpha$ and $\hat u$, respectively.
\end{pro}

As a consequence of this Proposition we conclude that $\hat u$ and
$\hat\alpha$ are new solutions of the MRKdV system (\ref{mr}). We
remark that, as usual, the Darboux transformation can be viewed as
a gauge transformation:
\begin{align*}
\psi&\to T_0\psi,\\
L&\to \hat L=T_0LT_0^{-1},\\
M&\to \hat M=\partial_tT_0\cdot T_0^{-1}+T_0MT_0^{-1},\\
T_0&:=D+\delta_0.
\end{align*}

To construct Crum type transformation,
let us start with $n$  solutions $\theta_i$,
$i =0,...,n-1$, of  equation (\ref{linear})
with eigenvalues as $\lambda=k_i$, $i=0,..., n-1$.
 To make sense, we choose
the $\theta_i $ in such way that its index indicates its
parity: those with even indices are even and with odd indices
are odd variables. We use $\theta_0$ to do our first
step  transformation and then $\theta_i$, $ i=1,\dots , n-1$, are transformed to new solutions $\theta_i[1]$
of the transformed linear equation and $\theta_0$
goes to zero. Next step can be effected by
using $\theta_1[1]$ to form a Darboux operator and at this
time $\theta_1[1]$ is lost. We can continue this
iteration process until all the seeds are mapped to zero. In
this way, we have
\begin{pro}
Let $\theta_i$, $i=0,\dots , n-1$, be solutions of
the linear system (\ref{linear}) with
$\lambda=k_i$, $i=0,\dots ,n-1$, and parities $ p(\theta_i)=(-1)^i$,
 then after $n$ iterations of the
Darboux transformation of Proposition 1, one obtains a new Lax
operator
\[
{\hat{L}}=T_nLT_{n}^{-1}, \quad T_n=D^n +\sum_{i=0}^{n-1}a_iD^i,
\]
where the coefficients $a_i$ of the gauge operator $T_n$ are defined
by
\begin{equation}\label{l2}
(D^n +\sum_{i=0}^{n-1}a_iD^i)\theta_j=0, \qquad j=0,\dots , n-1.
\end{equation}
\end{pro}

The explicit form of the transformed field variables is given by
${\hat{L}}=T_nLT_{n}^{-1}$ from where it follows that the new
fields $\hat\alpha$ and $\hat u$ can be written as
\begin{align*}
\hat{\alpha}&=(-1)^n\alpha-2\partial a_{n-1},\\
{\hat{u}}&=u-2\partial a_{n-2}-a_{n-1}((-1)^n \alpha +{\hat{\alpha}})+
\frac{1-(-1)^n}{2}D\alpha.
\end{align*}

Now, we must recall the reader that the Berezinian or superdeterminant
of an even matrix, say $\cal M=\big(\begin{smallmatrix}
 A & B\\ C &D
\end{smallmatrix}\big)$,   is
\[
\sdet\cal M=\frac{\det \left(A -B D^{-1}C\right)}
{\det D}=\frac{\det A}{\det\left(D- C A^{-1} B\right)}.
\]

To obtain the explicit expressions, we have to find
out the $a_{n-2}$ and $a_{n-2}$ by solving the linear system (\ref{l2}). 
In the even case $n=2k$ which is most interesting, we have
\[
a_{2k-2}=-\frac{\sdet\hat{\cal W}}{\sdet{\cal W}},\qquad
a_{2k-1}=D \ln\, \sdet {\cal W},
\]
where
\begin{align*}
{\boldsymbol a}^{(0)}:&=(a_0, a_2,\dots, a_{2k-2}),\qquad
{\boldsymbol a}^{(1)}:=(a_1, a_3,\dots, a_{2k-1}),\\
{\boldsymbol\theta}^{(0)}:&=(\theta_0,\theta_2,\dots,\theta_{2k-2}),
\qquad
{\boldsymbol\theta}^{(1)}:=(\theta_1,\theta_3,\dots,\theta_{2k-1}),
\\ {\boldsymbol
b}^{(i)}:&=\partial^k{\boldsymbol\theta}^{(i)},\qquad
W^{(i)}:=\begin{pmatrix}{\boldsymbol\theta}^{(i)}\\
\partial{\boldsymbol\theta}^{(i)}\\ \vdots\\ \partial^{k-1}
{\boldsymbol\theta}^{(i)}
\end{pmatrix},\qquad i=0,1,\\
\cal W&:=\begin{pmatrix}
W^{(0)}&W^{(1)}\\DW^{(0)}&DW^{(1)}\end{pmatrix},\quad
\hat{\cal W}:=\left(\begin{array}{cc} {\hat W}^{(0)}&{\hat W}^{(1)}\\
DW^{(0)} &DW^{(1)}\end{array}\right)
\end{align*}
 and $ {\hat{W}}^{(0)} $ and $ {\hat{W}}^{(1)}$ are obtained from the matrices
$W^{(0)}$ and $W^{(1)}$ by replacing the last rows with
${\boldsymbol b}^{(0)}$ and ${\boldsymbol b}^{(1)}$, respectively.
It should be noticed that the supermatrix $\cal W$ is even and  has
a Wronski type structure.

Summarizing the above results, we now have the following
\begin{th}
Let $\alpha,u$ be a  solution of (\ref{mr}) and
$\{\theta_j\}_{j=0}^{n-1}$ be a set of $n(=2k)$ solutions of the
associated linear system (\ref{linear}), such that the parity is
$p(\theta_j)=(-1)^j$. Then,  we have  new solutions $\hat\alpha$,
$\hat u$ of (\ref{mr}) given by
\begin{align*}
\hat\alpha&=\alpha-2 D^3\ln\,\sdet{\cal W},\\
\hat u&=u+2\partial\Big(\frac{\sdet\hat{\cal W}}{\sdet {\cal W}}\Big)+
(\alpha+\hat\alpha) D\ln\,\sdet{\cal W}.
\end{align*}
\end{th}

Let us remark that our Darboux transformations are useful even outside the 
MRKdV system. Indeed, the most obvious application is to the
SUSY KP equation (\cite{ueno}) which is a closed system obtained from
supersymmetric KP hierarchy. Since the Lax pair is essentially
the one we had for MRKdV, our Darboux transformations can be
used directly in this context.

Solutions of the MRKdV are  found rarely, see for example
\cite{radul,ibort}. In \cite{lm1} and \cite{lm2} one can find examples
of solutions, in particular in \cite{lm2} we gave an interesting solution
which behaves like a genuine supersoliton.

Our solution is
\[
{\hat \alpha}=-2\partial a_{1}, \qquad 
{\hat u}=-2\partial a_{0}.
\]
with
\begin{align*}
a_0&=f-k\big(\gamma^{(1)}_-\gamma^{(1)}_+
+\vartheta(\gamma^{(1)}_+\gamma^{(0)}_--\gamma^{(1)}_-
\gamma^{(0)}_+)\big)g-
\vartheta k\big(c^{(0)}_+\gamma^{(1)}_--c^{(0)}_-\gamma^{(1)}_+\big)fg,\\
a_1&=\big(k(c^{(0)}_+\gamma^{(1)}_--c^{(0)}_-\gamma^{(1)}_+)+
\vartheta(c^{(0)}_+\gamma^{(0)}_--c^{(0)}_-
\gamma^{(0)}_+)\big)g,
\end{align*}
where
\begin{align*}
f&:=-k\left(
\frac{c_{+}^{(0)}\exp(\eta)-c_{-}^{(0)}\exp(-\eta)}
{c_{+}^{(0)}\exp(\eta)+c_{-}^{(0)}\exp(-\eta)}
\right),\\
g&:=\frac{2}{\left(c_{+}^{(0)}\exp(\eta)+c_{-}^{(0)}\exp(-\eta)\right)
\left(\gamma_{+}^{(0)}\exp(\eta)+\gamma_{-}^{(0)}\exp(-\eta)\right)}.
\end{align*}
here $\eta=k x+k^{3}t$, $k\in\Bbb C$ and
 $c_{\pm}^{(i)}$, $\gamma_{\pm}^{(i)}$ are supernumbers with
parities indicated by the superfix.

Notice that our solution can be understood as a supersoliton
which has the KdV soliton, $-2\partial f$, as its body, and that
the choice $c_{+}^{(0)}=\gamma_{+}^{(0)}$ and 
$c_{-}^{(0)}=\gamma_{-}^{(0)}$ gives the solution found in \cite{ibort}.

\leavevmode

\epsfxsize=6cm

\hspace*{3cm}\epsffile{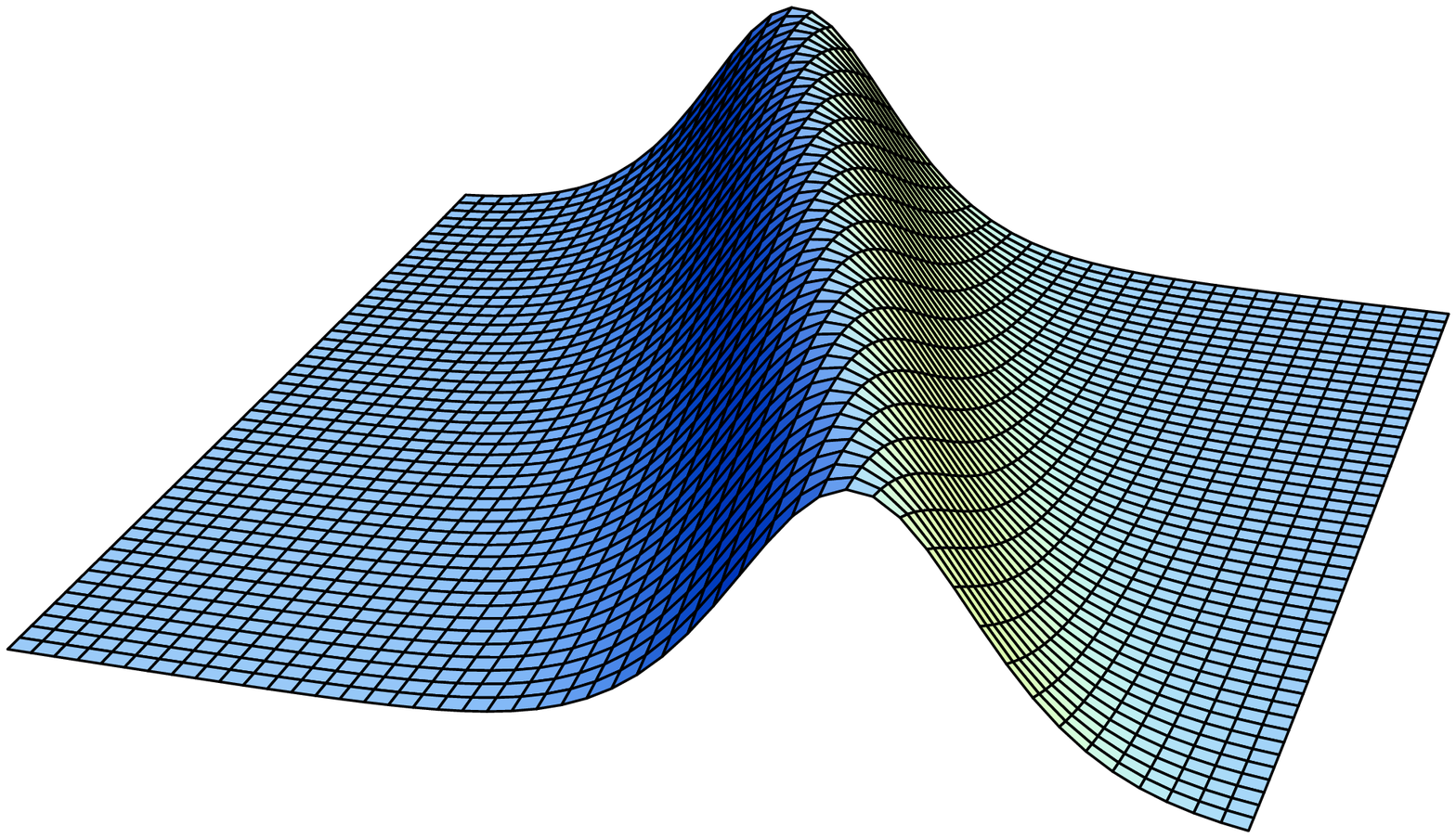}

\epsfxsize=6cm

\hspace*{3cm}\epsffile{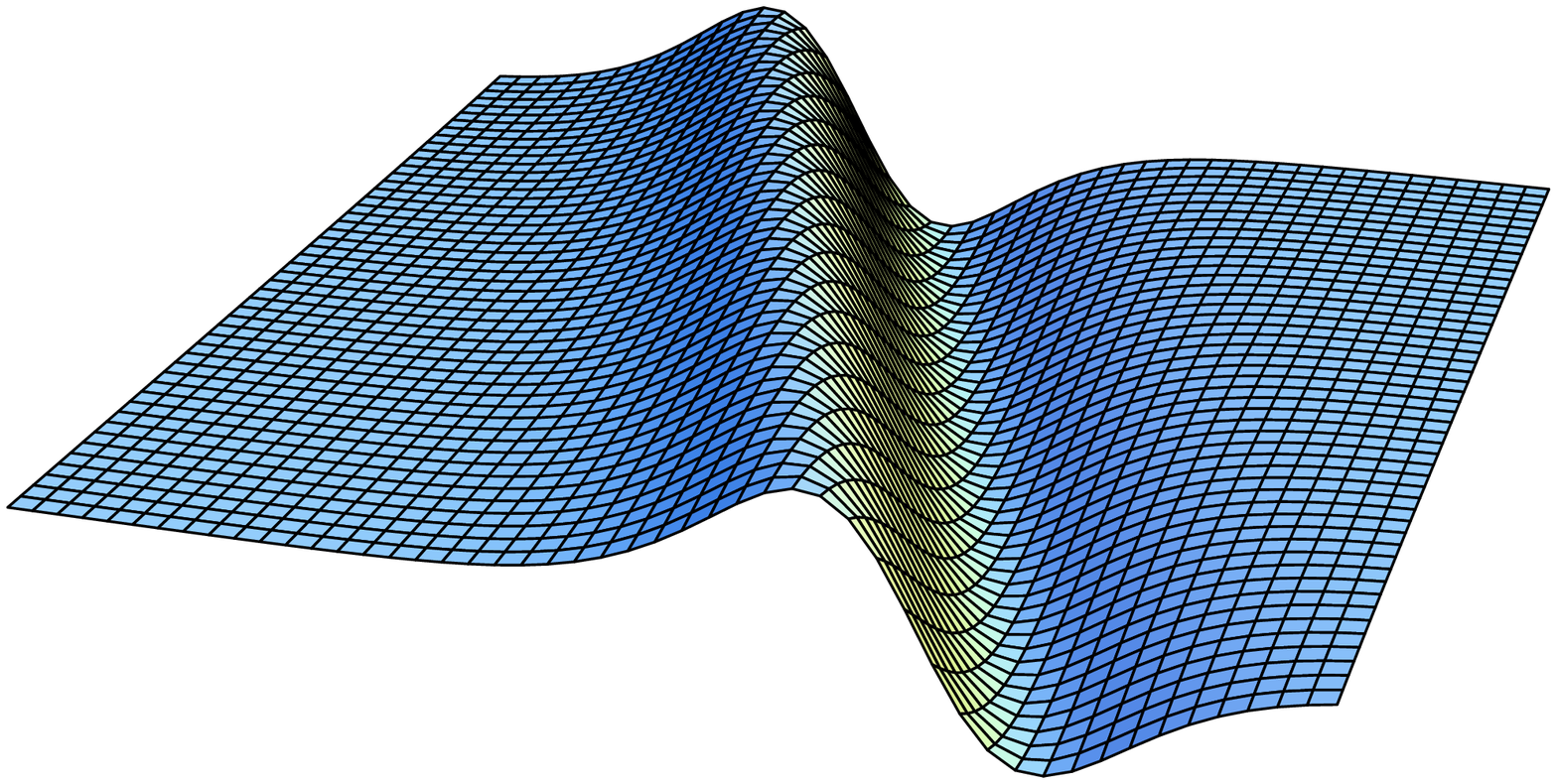}

\vspace*{.5cm}

\epsfxsize=6cm
\hspace*{3cm}\epsffile{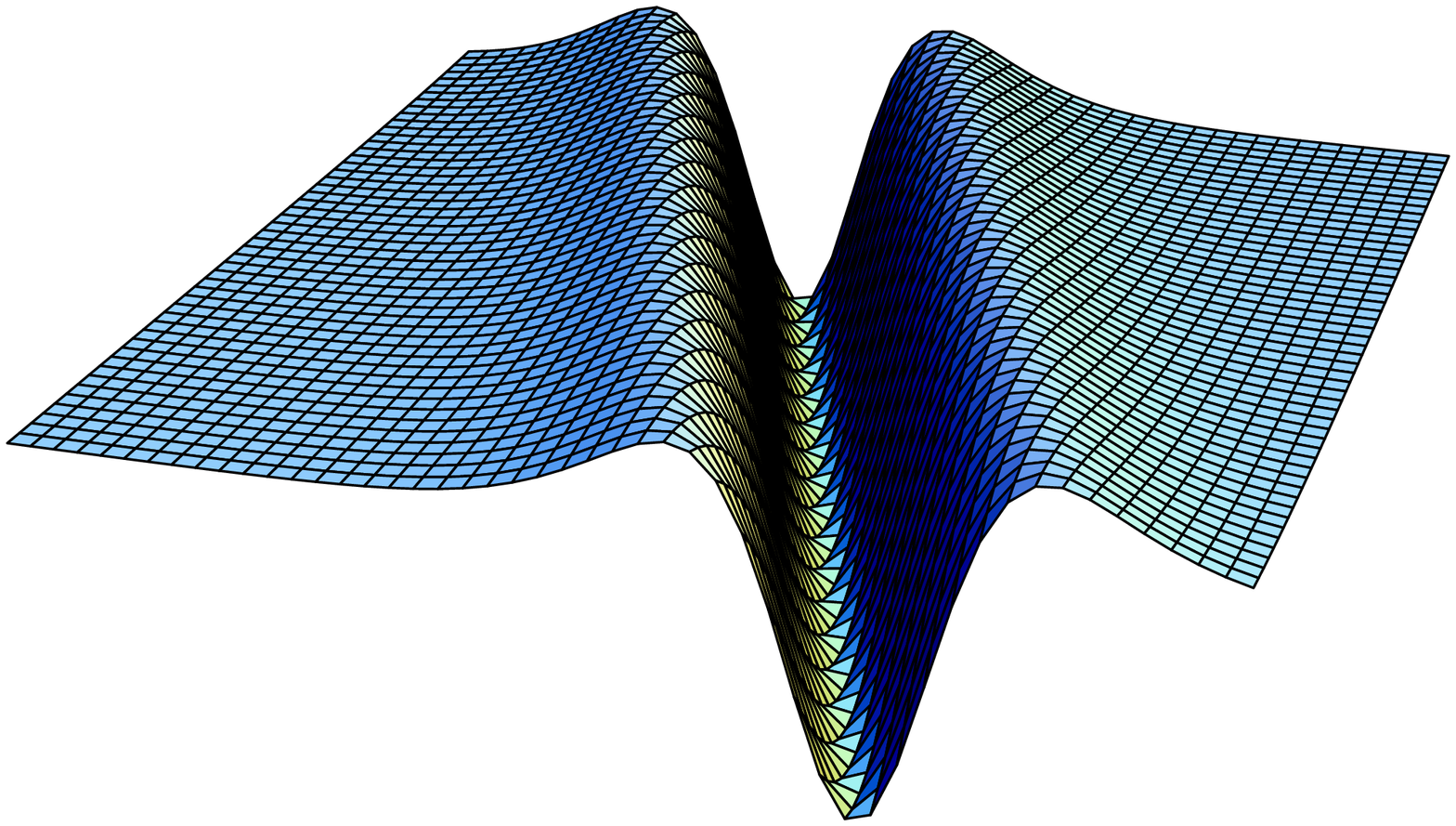}

In the figures we plot, in the real $x$-$t$ plane, 
the functions $f_x$, $g_x$ and $(fg)_x$ that appear in the construction
of the solution corresponding to the data:
$k=1.1$, $c_+^{(0)}=1$, $c_-^{(0)}=1.5$, $\gamma_+^{(0)}=1.2$ and
$\gamma_+^{(0)}=2$.

\section{Darboux Transformations for the Mathieu KdV and
SUSY Sine-Gordon Equations}

The  Mathieu KdV equation reads (\cite{m-r,ma2})
\[
\alpha_t={1\over 4}(\alpha_{xx}+3\alpha D\alpha)_x,
\]
which is obtained from (\ref{mr}) by setting $u=0$, being
the Lax operator
\[
L=\partial^2+\alpha D.
\]
In \cite{liu} one of the authors presented
a preliminary version of 

\begin{pro}
If
\begin{equation}
\psi_{xx}+\alpha D\psi=\lambda \psi, \qquad
\end{equation}
and $\psi_0$ is an even solution of above equation with $\lambda=
\lambda_0$ such that the constants
\begin{align*}
J(\psi_0)=&\psi_{0,x}^{2}+2(D\psi_0)_xD\psi_0-\lambda_0\psi_0^2,\\
I(\psi_0,\psi)=&(\lambda-\lambda_0)\left(D^{-1}((D\psi_0)\psi)\right)+\lambda_0\psi_0\psi-
\psi_{0,x}\psi_x\\ &-(D\psi_0)_xD\psi+(D\psi_0)D\psi_x
\end{align*}
vanish, then
\[
{\hat\psi}:=\psi_0^{-1} D^{-1}(\psi_0 D\psi-(D\psi_0)\psi),\quad
{\hat\alpha}:=\alpha-4D^3 \ln \psi_0
\]
satisfy
\[
{\hat\psi}_{xx}+\hat{\alpha} D{\hat\psi}=\lambda {\hat\psi}, \qquad
\]
\end{pro}

Compared with the result of \cite{liu}, the above Proposition is 
an improved version in the sense that the constant $I(\psi_0,\psi)$
has a much simpler structure. 
An application of this Darboux transformation
is to the supersymmetric sine-Gordon system, which
reads (\cite{ik})
\begin{equation}\label{sg}
DD_t\Phi=2\cosh (2\Phi)
\end{equation}
where $D_t={\partial\over\partial \theta_t}+\theta_t{\partial\over
\partial t}$ and $\theta_t$ is another Grassmann odd variable.

According to \cite{ik}, the linear problems are
\begin{align*}
D\psi_1+\lambda\psi_3+\psi_4&=0, \\
D\psi_2-\lambda\psi_3+\psi_4&=0,\\
D\psi_3+2(D\Phi)\psi_3+\psi_1+\psi_2&=0,\\
D\psi_4-2(D\Phi)\psi_4+\lambda(\psi_1-\psi_2)&=0,
\end{align*}
and
\begin{align*}
D_t\psi_1+\exp(2\Phi)\psi_3-\lambda^{-1}\exp(-2\Phi)\psi_4&=0,\\
D_t\psi_2+\exp(2\Phi)\psi_3+\lambda^{-1}\exp(-2\Phi)\psi_4&=0,\\
D_t\psi_3+\lambda^{-1}\exp(-2\Phi)(\psi_1-\psi_2)&=0,\\
D_t\psi_4-\lambda^{-1}\exp(2\Phi)(\psi_1+\psi_2)&=0.
\end{align*}

Introducing ${\tilde \psi}_1:=\psi_1-\psi_2$ and
${\tilde \psi}_2:=\psi_1+\psi_2$, we easily see that the above
linear system can be written as follows:
\begin{equation}\label{sg1}
\begin{pmatrix}\tilde \psi_1\\{\tilde \psi}_2\end{pmatrix}_x
=2\begin{pmatrix}-(D\Phi)D&\lambda\\
\lambda&(D\Phi)D\end{pmatrix}
\begin{pmatrix}\tilde \psi_1\\{\tilde \psi}_2\end{pmatrix}
\end{equation}
and
\begin{equation}\label{sg2}
D_t\begin{pmatrix}\tilde \psi_1\\{\tilde \psi}_2\end{pmatrix}
=\begin{pmatrix}0&-\lambda^{-1}\exp(-2\Phi)D\\
\lambda^{-1}\exp(2\Phi)D&0\end{pmatrix}
\begin{pmatrix}\tilde \psi_1\\{\tilde \psi}_2\end{pmatrix}.
\end{equation}

Differentiating (\ref{sg1}) we get
\begin{align*}
{\tilde \psi}_{1,xx}&=-\alpha_1 D{\tilde\psi}_1+4\lambda^2{\tilde\psi}_1,\\
{\tilde\psi}_{2,xx}&=\alpha_2 D{\tilde\psi}_2+4\lambda^2{\tilde\psi}_2
\end{align*}
where  $\alpha_1:=\gamma_x -\gamma D\gamma$ and
$\alpha_2:=-\gamma_x -\gamma D\gamma$ with $\gamma=2D\Phi$.

Now a slight modification of the previous Darboux transformation
yields
\begin{th}
If $\Phi$ is a solution of the SUSY sine-Gordon equation (\ref{sg})
and $\tilde\psi_1$ and $\tilde\psi_2$ are particular solutions
of  (\ref{sg1}) and (\ref{sg2}) with spectral parameter
$\lambda=\lambda_0$, such that $J(\tilde\psi_1)=J(\tilde\psi_2)=0$, then
\[
{\hat \Phi}=\Phi+\ln \frac{\tilde\psi_1}{\tilde\psi_2},
\]
is a new solution of (\ref{sg}).
\end{th}

\noindent
{\em Remarks}
\begin{enumerate}
\item The method we used is a generalization of the one by
 \cite{wadati} for the classical sine-Gordon equation. We also notice that
 this idea was used in \cite{nimmo} for the two-dimensional sine-Gordon
 equation of Konopelchenko and Rogers.
\item We may also obtain Darboux type transformations for
 the SUSY modified KdV equation (\cite{ma2}).
\item We conjecture that  Crum type
iteration of the above Darboux transformation
will be represented in
terms of Pfaffians instead of Wronskians.
\end{enumerate}

The application of this theorem to the most simple case, namely
$\Phi=\I\pi/4$ yields the following interesting SUSY extension
of the kink solution. Namely, we have
the following solution
\[
\Phi=\frac{\I\pi}{4}+\ln\frac{1+{\cal A}\exp(-4\eta)}{1-{\cal A}\exp(-4\eta)}
\]
where
\begin{align*}
{\cal A}(\theta,\theta_t)&:=
\frac{A_0(1-2\I\theta\theta_t)+
A_1(\theta+\I\lambda^{-1}\theta_t)}{
c^{-1}A_0(1-2\I\theta\theta_t)+c
A_1(\theta-\I\lambda^{-1}\theta_t)},\\
\eta(x,t)&:=\lambda x+\lambda^{-1}t,
\end{align*}
with $A_0$ and $c$  even supernumbers
with nonvanishing body, $\lambda\in\C$
and $A_1$ an odd supernumber.

This solution can be thought as a superkink, in fact its body is
\[
\Phi_{\text{body}}=
\frac{\I\pi}{4}+\ln\frac{1+c_{\text{body}}\exp(-4\eta)}
{1-c_{\text{body}}\exp(-4\eta)},
\]
that after taking the appropriate Wick rotation goes into the
standard kink solution of the sine-Gordon equation.
Obviously, the soul of this solution is far from trivial.

{\bf Acknowledgement}.
The present paper is the extended version of the talk presented at 
the meeting by one of us (MM). A number of the results were
obtained during or after the meeting. MM would like to thank
the organizers for the hospitality and coverage of local expenses;
and also {\em Bolsa de Viaje  Complutense 1997}.

\end{document}